\documentclass[aps,epsf,preprint]{revtex4}
\usepackage{latexsym}
\usepackage{amsfonts}
\usepackage{graphicx}
\usepackage{epsfig}

\begin{document}

\begin{center}

{\large\large\bf Space-Time Correlations in the Orientational Order
Parameter and the Orientational Entropy of Water}

\bigskip

Pradeep Kumar,$^1$ Sergey V. Buldyrev,$^2$ and H. Eugene Stanley$^3$

\bigskip

{\scriptsize

$^1$ Center for Studies in Physics and Biology

The Rockefeller University, 1230 York Avenue

New York, NY 10021 USA

$^2$Department of Physics, Yeshiva University

500 West 185th Street, New York, NY 10033 USA

$^3$Center for Polymer Studies and Department of Physics

Boston University, Boston, MA 02215 USA

~~

\bigskip

(kbs22jul.tex --- Revised 22 July 2008)}

\bigskip
\bigskip

{\scriptsize ABSTRACT}

\end{center}

\noindent We introduce the spatial correlation function $C_Q(r)$ and
temporal autocorrelation function $C_Q(t)$ of the local tetrahedral
order parameter $Q\equiv Q(r,t)$.  Using computer simulations of the
TIP5P model of water, we investigate $C_Q(r)$ in a broad region of the
phase diagram. First we show that $C_Q(r)$ displays anticorrelation at
$r\approx 0.32$nm at high temperatures $T>T_W\approx 250$~K, which
changes to positive correlation below the Widom line $T_W$. Further we
find that at low temperatures $C_Q(t)$ exhibits a two-step temporal
decay similar to the self intermediate scattering function, and that the
corresponding correlation time $\tau_Q$ displays a dynamic crossover
from non-Arrhenius behavior for $T>T_W$ to Arrhenius behavior for
$T<T_W$. Finally, we define an orientational entropy $S_Q$ associated
with the {\it local\/} orientational order of water molecules, and show
that $\tau_Q$ can be extracted from $S_Q$ using an analog of the
Adam-Gibbs relation.

\bigskip\bigskip\bigskip\bigskip

\noindent
The local structure around a water molecule arising from the vertices
formed by four nearest neighbors is approximately {\it tetrahedral}. The
degree of tetrahedrality \cite{soper,sato,tanaka} can be quantified by
the local tetrahedral order parameter $Q$
\cite{errington,errington02,yan05,Errington06,steinhardt}. Spatial
fluctuations in $Q$ suggest a high degree of orientational
heterogeneity. The distribution of $Q$ is bimodal, centered around less
tetrahedral and more tetrahedral values of $Q$. Upon decreasing
temperature, the high-tetrahedrality peak grows, suggesting that the
local structure of water becomes much more tetrahedral at low enough
temperatures \cite{kumarPRL06,kumarPNAS07}.

Water has been hypothesized to belong to the class of polymorphic
liquids, phase separating---at sufficiently low temperatures and high
pressures---into two distinct liquid phases: a high density liquid (HDL)
with smaller $Q$ and a low density liquid (LDL) with larger $Q$
\cite{Poole1}.  The co-existence line separating these two phases
terminates at a liquid-liquid (LL) critical point.  The locus of maximum
correlation length in the one-phase region is called the Widom line $T_W
\equiv T_W(P)$, near which different response functions such as isobaric
heat capacity $C_P$ and isothermal compressibility $K_T$ display maxima.
Recent neutron scattering experiments \cite{Liu04} and computer
simulations \cite{xuPNAS} show that the dynamics of water changes from
non-Arrhenius for $T>T_W$ to Arrhenius for $T< T_W$.

In this paper, we ask how the orientational order and its
spatio-temporal correlations change upon crossing the Widom line. To
this end, we introduce the spatial correlation function $C_Q(r)$ and the
temporal autocorrelation function $C_Q(t)$ for the local orientational
order of water, which can also be applied to study other locally
tetrahedral liquids such as silicon~\cite{sastry2003,Morishita06},
silica~\cite{voivod2000}, phosphorus~\cite{katayama2004}; all of which
have also been found to display some of the water-like anomalies. To
this end, we perform molecular dynamics (MD) simulations of $N=512$
waterlike molecules interacting via the TIP5P potential
\cite{JorgensenXX,YamadaXX}, which exhibits a liquid-liquid (LL)
critical point at $T_C\approx 217$~K and $P_C\approx 340$~MPa
\cite{YamadaXX,Paschek05}.  We carry out simulations in the $NPT$
ensemble at atmospheric pressure ($P=1$~atm) for temperatures $T$
ranging from $320$~K down to $220$~K.

To quantify the local degree of orientational order, we calculate the
local tetrahedral order parameter \cite{errington}
\begin{equation}
Q_k \equiv 1-\frac{3}{8}\sum_{i}^{3}\sum_{j=i+1}^{4}
\left[\cos\psi_{ikj}+\frac{1}{3}\right]^2,
\label{eq:qk}
\end{equation}
where $\psi_{ikj}$ is the angle formed by the molecule $k$ and its
nearest neighbors $i$, and $j$. The average value $\langle
Q\rangle\equiv (1/N)\sum_kQ_k$ increases with decreasing $T$, and
saturates at lower $T$, while $|d\langle Q\rangle/dT|$ has a maximum at
the Widom line $T_W\approx 250$~K \cite{kumarPNAS07,kumarPRL06}.

To characterize the spatial correlations of the local order parameter we
find all the pairs of molecules $i$ and $j$ whose oxygens are separated
by distances belonging to the interval $[r-\Delta r/2,r+\Delta
 r/2]$. The number of such pairs is
\begin{equation}
N(r,\Delta r)=\sum_{ij}\delta(r_{ij}-r,\Delta r),
\end{equation}
where $r_{ij}$ is the distance between the oxygens of molecules $i$ and
$j$. The sum is taken over all molecules in the system and
\begin{equation}
\delta(r_{ij}-r,\Delta r)=\cases{1 & if $|r-r_{ij}|<\Delta r/2$\cr
                              0 & otherwise.}
\end{equation}
For $\Delta r\to 0$, $\delta(r_{ij}-r,\Delta
r)\to\delta(r_{ij}-r)\Delta r$ where $\delta(r_{ij}-r)$ is the Dirac
$\delta$-function. $N(r,\Delta r)$ can be approximated as
\begin{equation}
N(r,\Delta r)=4\pi r^2Ng_{OO}(r)\rho\Delta r.
\end{equation}
where $N$ is the total number of molecules in the system, $g_{\rm
 OO}(r)$ is the oxygen-oxygen pair correlation function, and $\rho$ is
the number density.

Next we find the local order parameters $Q_i$ and $Q_j$ for all such
pairs of molecules and compute their mean
\begin{equation}
Q(r) \equiv \langle Q\rangle_r\equiv
     {\sum_{ij}Q_i\delta(r_{ij}-r,\Delta r)\over N(r_{ij}-r,\Delta
       r)},
\end{equation}
their variance
\begin{equation}
\sigma_Q^2(r)\equiv\langle
Q^2\rangle_r - \langle Q \rangle_r^2 \equiv \sum_{ij}{Q_i^2\delta(r_{ij}-r,\Delta r)\over
 N(r_{ij}-r,\Delta r)} - \langle Q \rangle_r^2,
\end{equation}
and covariance
\begin{equation}
\langle Q_iQ_j\rangle_r\equiv \sum_{ij}{Q_{i}Q_j\delta(r_{ij}-r,\Delta r)\over
 N(r_{ij}-r,\Delta r)}.
\end{equation}
Finally we introduce the spatial correlation function of the local
tetrahedrality $Q$
\begin{equation}
C_Q(r)\equiv {\langle Q_iQ_j\rangle_r-\langle Q\rangle_r\langle
 Q\rangle_r\over \sigma_Q^2(r)}.
\end{equation}

\begin{figure*}[htb]
\begin{center}
\includegraphics[width=7cm]{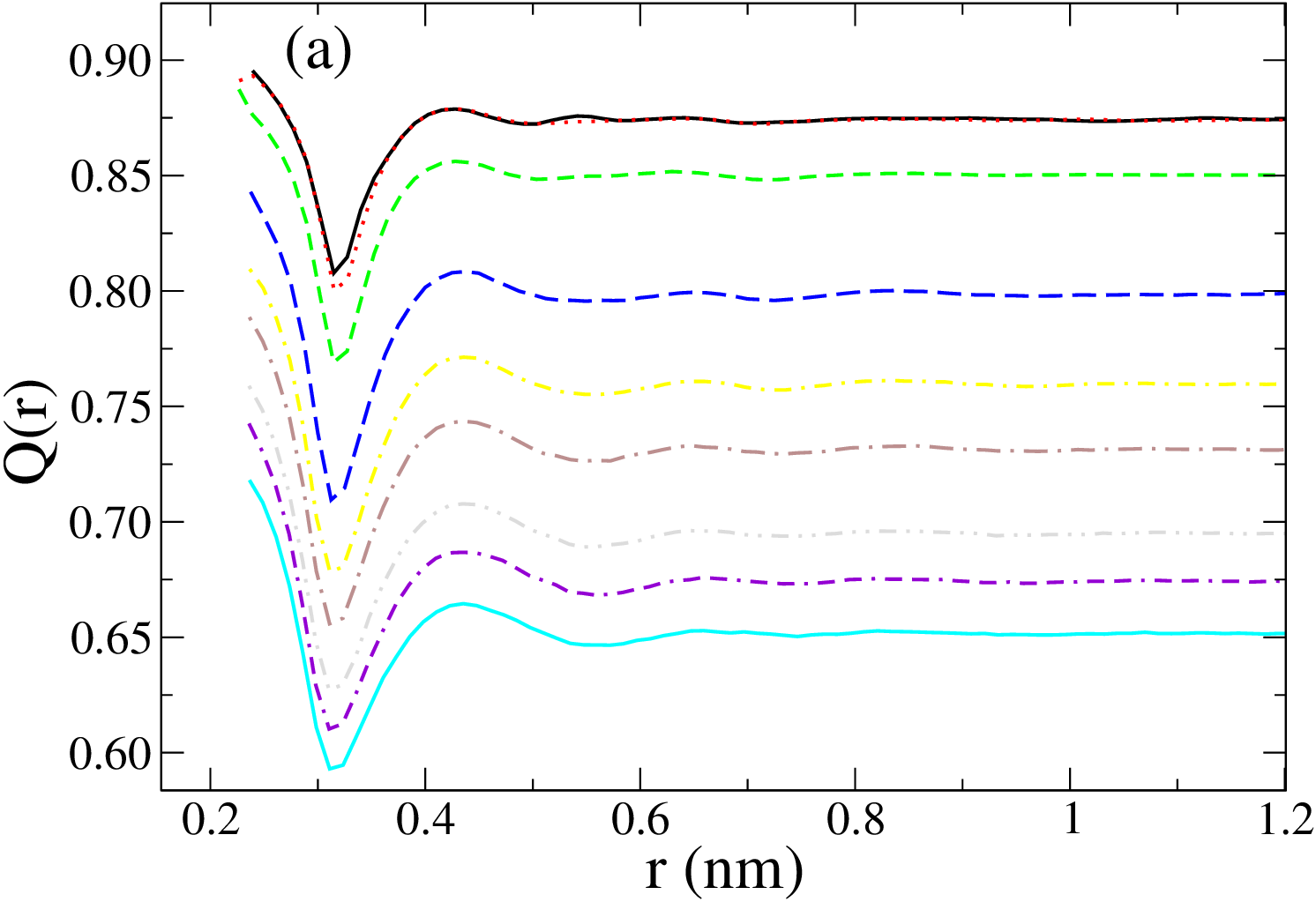}
\includegraphics[width=7cm]{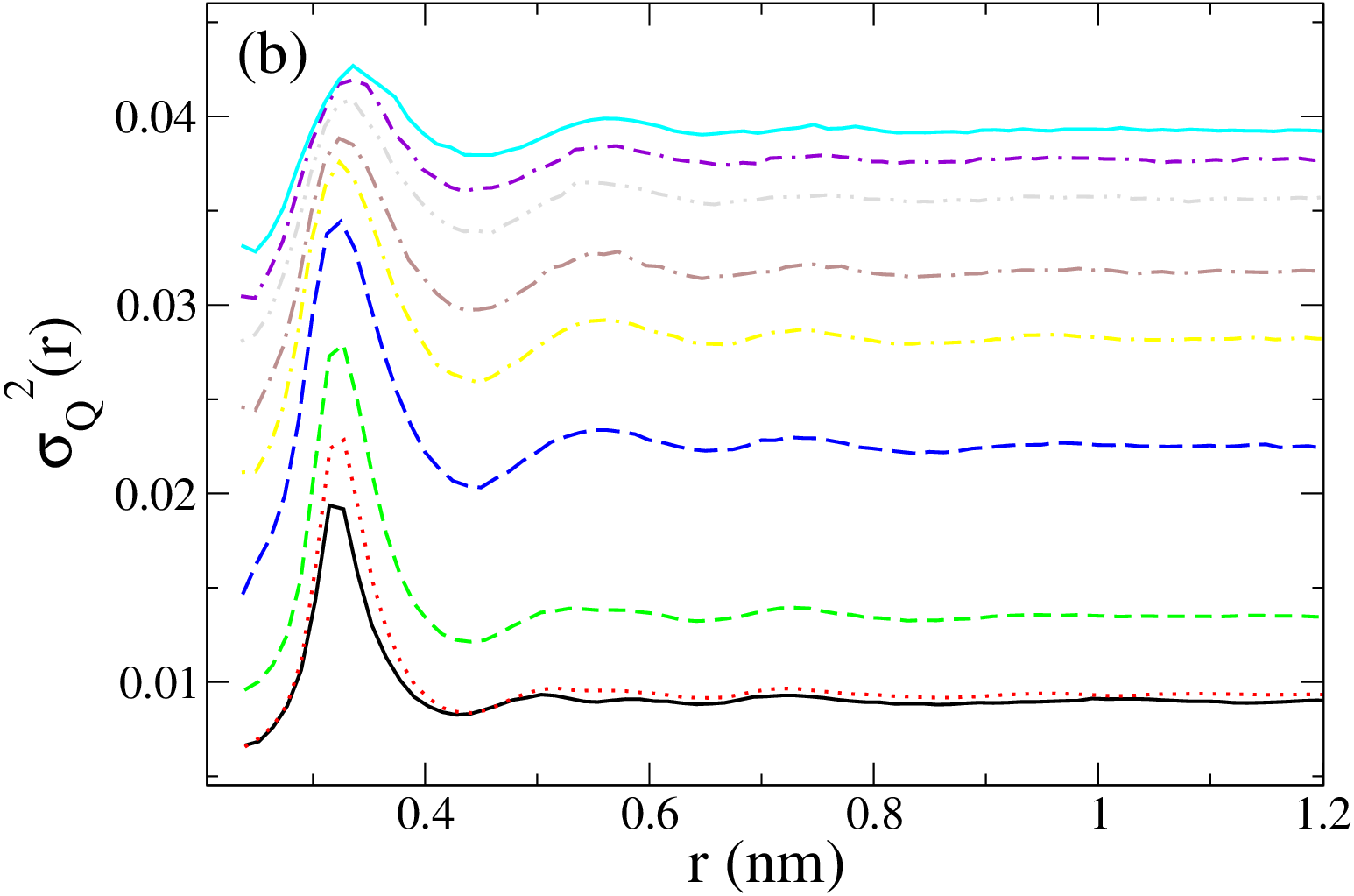}
\includegraphics[width=7cm]{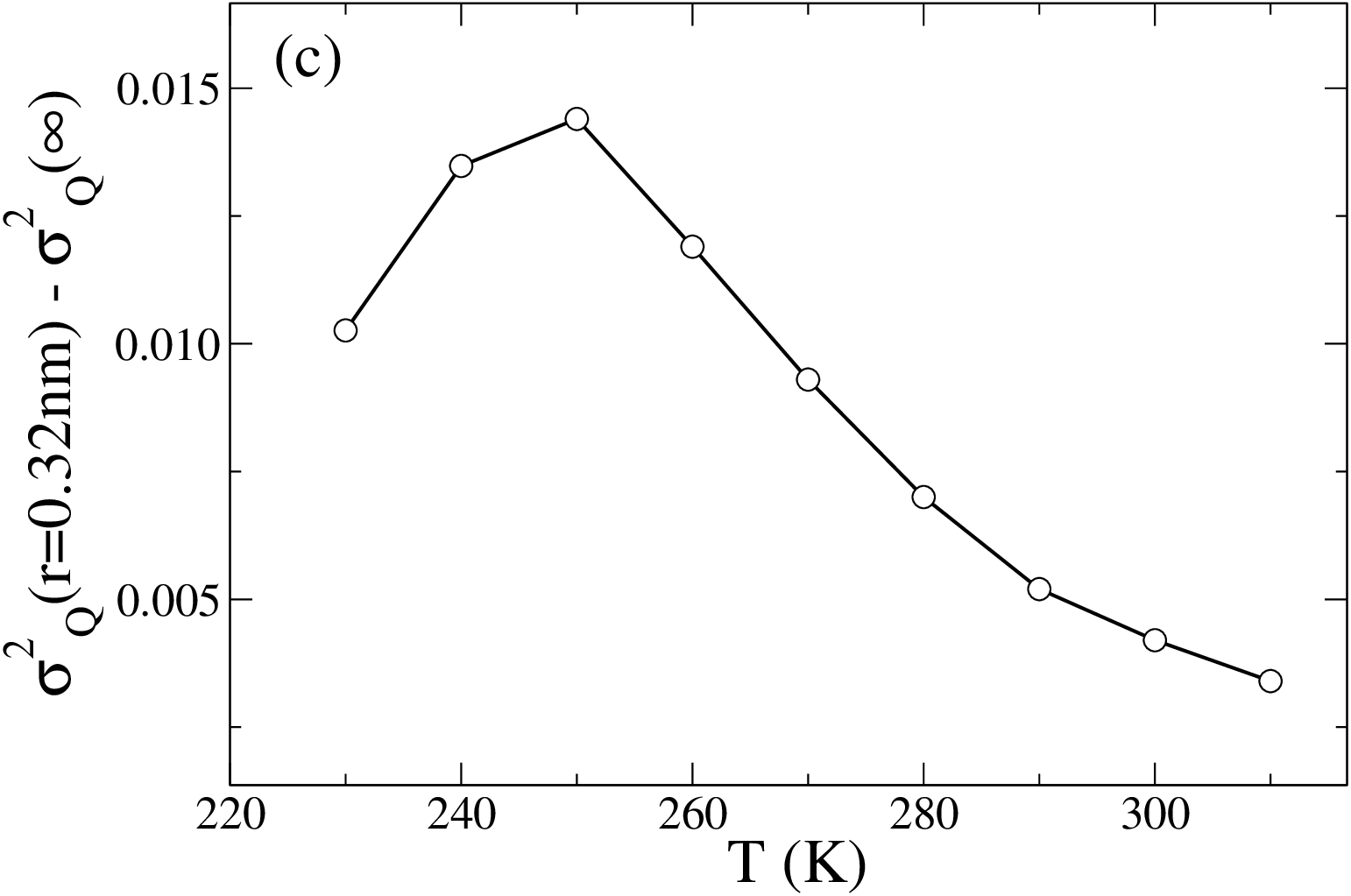}
\includegraphics[width=7cm]{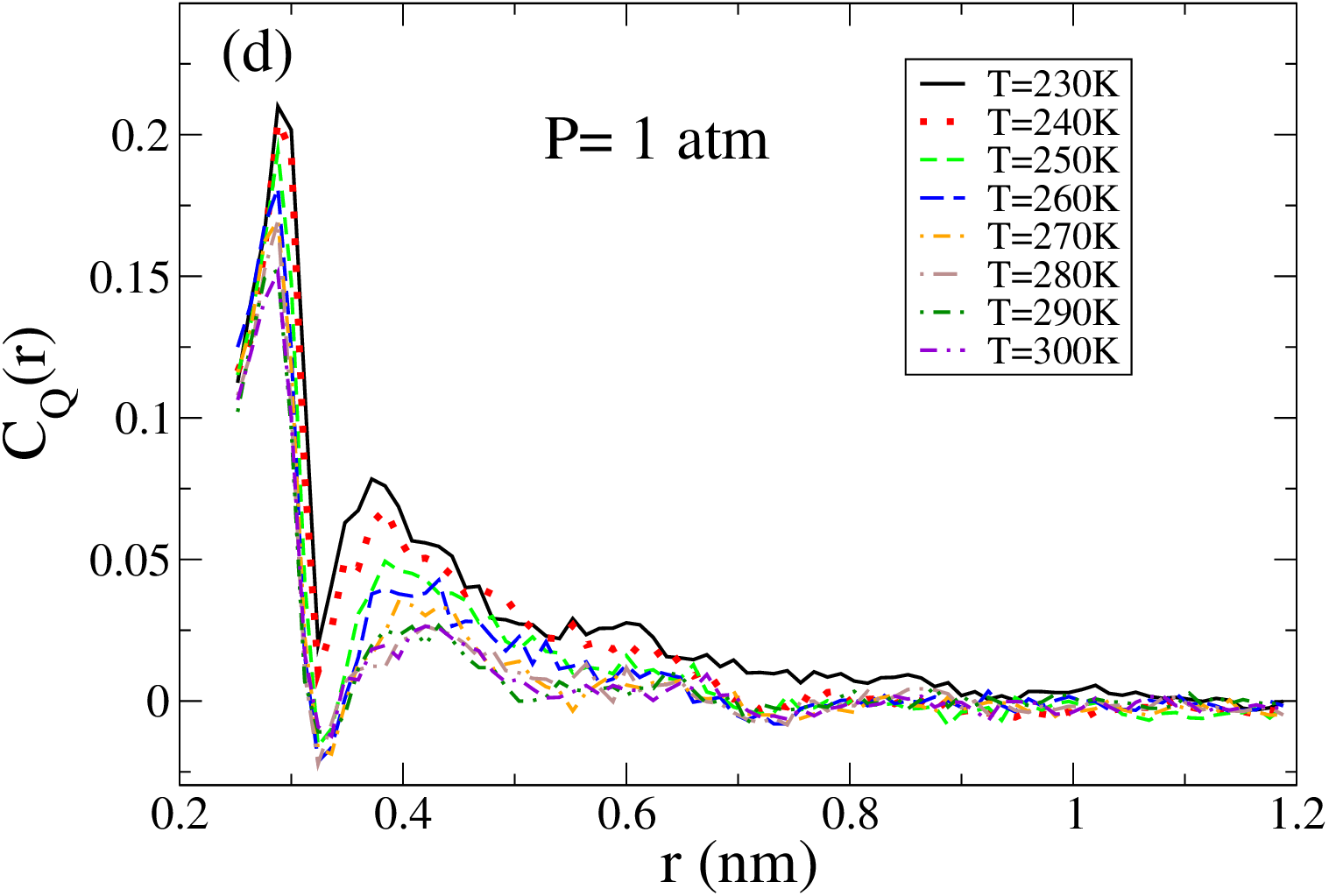}
\end{center}
\caption{(color online) (a) The average order parameter $Q(r)$ and (b)
 its variance, $\sigma_Q^2(r)$, as a function of the distance $r$. (c)
 Temperature dependence of $\sigma_Q^2(r=0.32$nm$)-\sigma_Q^2(\infty)$
 shows a maximum at the Widom temperature, suggesting the local
 fluctuations in $Q$ at the fifth-neighbor distance increases upon
 decreasing temperature and has a maximum at the Widom line. (c)
 Spatial correlation function $C_Q(r)$ of orientational order
 parameter $Q$ [Eq.~(\ref{eq:qk})], at various temperatures for
 pressures $P=1$~atm. $C_Q(r)$ has positive peaks at the positions of
 the nearest neighbor peaks in oxygen-oxygen pair correlation function
 $g_{\rm OO}(r)$. A negative minimum at the fifth neighbor distance
 $r\approx 0.32$~nm for high $T$ implies that the local tetrahedral
 order of a central molecule and its fifth neighbor are anticorrelated
 at $T>250$K. Interestingly the anticorrelation at $r=0.32$~nm changes
 to positive correlations upon crossing below the Widom line around
 $T_W\approx 250$~K.}
\label{fig:qpmr}
\end{figure*}

In Figs.~\ref{fig:qpmr}(a) and \ref{fig:qpmr}(b), we show $Q(r)$ and
$\sigma_Q^2(r)$ for different temperatures and atmospheric pressure. The
behavior of $Q(r)$ and its variance $\sigma_Q^2(r)$ as a function of the
distance $r$ has a clear physical meaning. For the molecules separated
by $0.32$~nm, $Q(r)$ has a deep minimum characterizing the distortion of
the first tetrahedral coordination shell of the four nearest neighbors
by the intrusion of a ``fifth neighbor''
\cite{sciortino1990}. Conversely, the quantity
$\sigma_Q^2(r=0.32$nm$)-\sigma_Q^2(\infty)$ for such molecules
dramatically increases upon decreasing temperature [see
 Fig.~\ref{fig:qpmr}(c)], and has a maximum at $T \approx 246$~K which
is approximately equal to the ambiant pressure value of $T_W$ ($\approx
250$~K for the TIP5P model) reported in Refs.~\cite{xuPNAS,kumarPNAS07}.
In Fig.~\ref{fig:qpmr}(d), we show $C_Q(r)$ for different temperatures
for $P=1$~atm. $C_Q(r)$ has positive maxima at positions of the first
and second peaks in the oxygen-oxygen pair correlation function,
suggesting that the molecules located at the tetrahedral positions are
strongly correlated in their $Q$ values. Water molecules separated by
$r\approx 0.32$nm exhibit weak anticorrelation in local tetrahedral
order at high temperatures, which changes to positive correlations upon
decreasing temperature below $T_W$.

\begin{figure}
\begin{center}
\includegraphics[width=8cm]{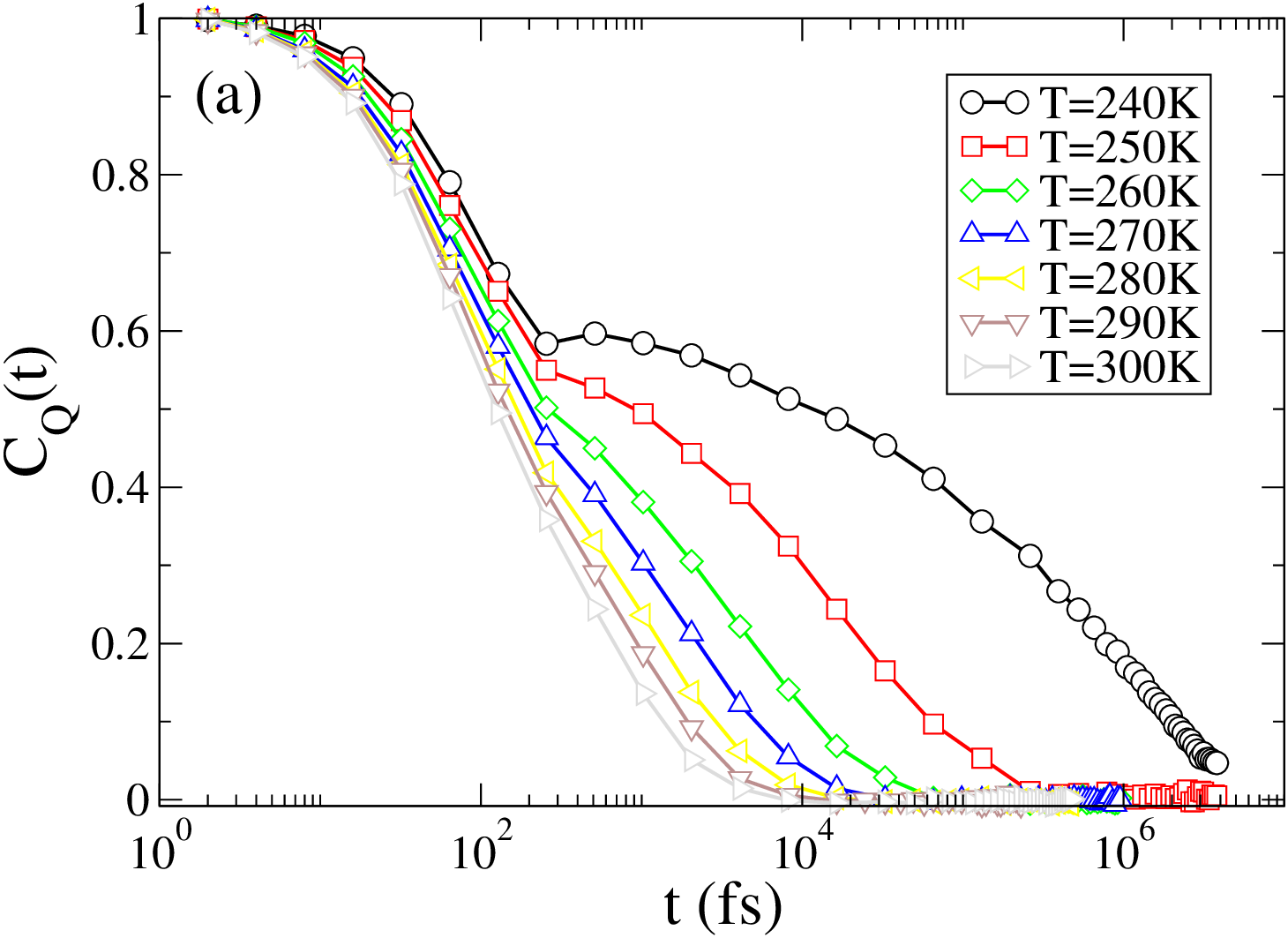}
\includegraphics[width=8cm]{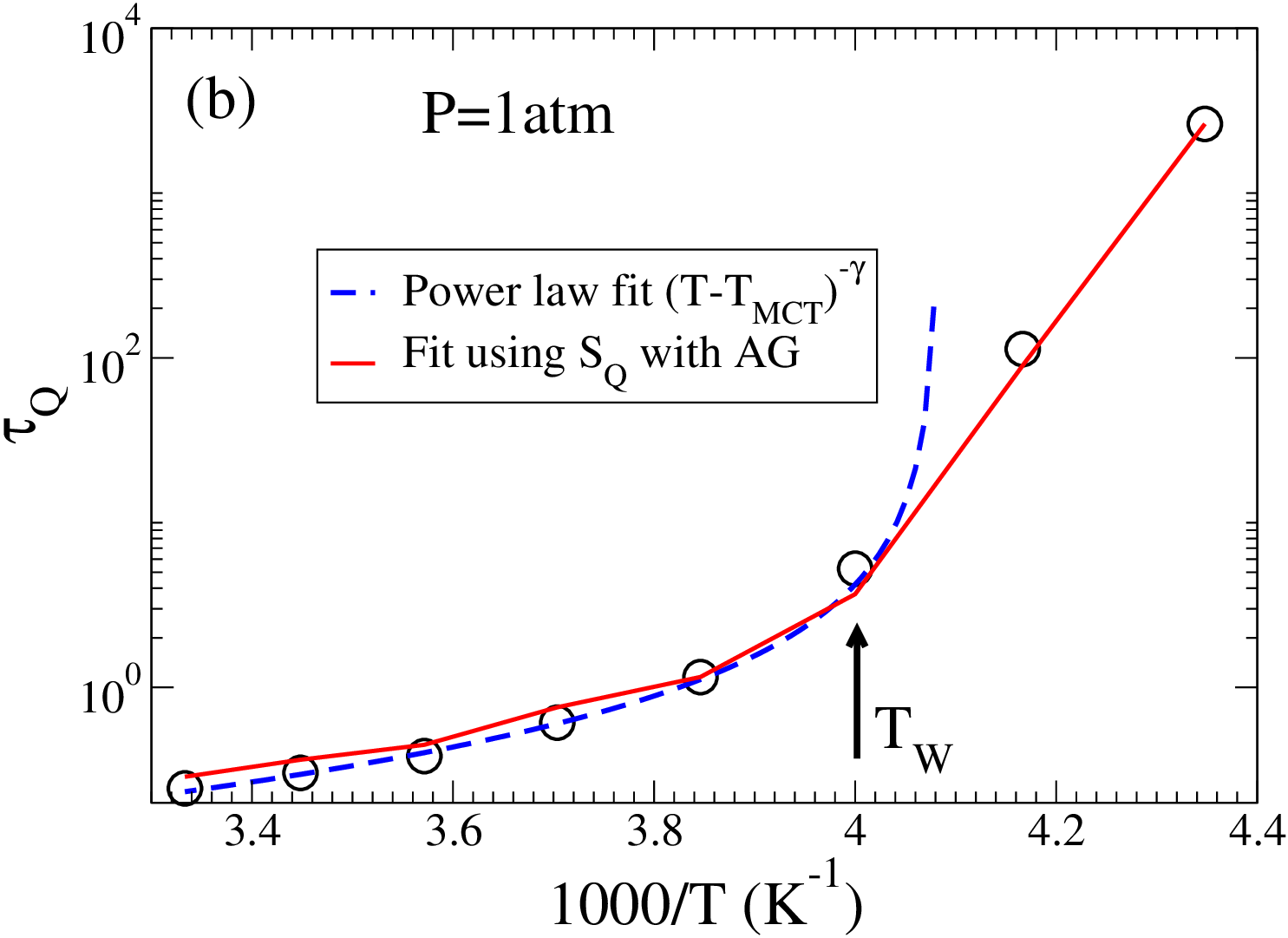}
\end{center}
\caption{(color online) (a) Autocorrelation function $C_Q(t)$ of
 orientational order parameter $Q$ at various temperatures. $C_Q(t)$ is
 exponential at high temperatures but displays a visible two step decay
 at low temperatures. (b) Correlation time $\tau_Q$ extracted from
 $C_Q(t)$. Solid line is the fit using the Adam-Gibbs relation
 [Eq.~(\ref{eq:ag})] between the orientational entropy $S_Q(T)$, and
 the orientational relaxation time $\tau_Q$. The dotted lines in (b) is
 the power law fit $(T-T_{\rm MCT})^{-\gamma}$. Behavior of $\tau_Q$
 deviates from the power law fit and has a crossover to Arrhenius
 behavior at low temperature near the same temperature $T_W$ where the
 response functions have maxima, which we identify with crossing the
 Widom line $T_W$.}
\label{fig:qpm-time}
\end{figure}

To study the time development of local tetrahedral orientational
order parameter, we introduce the time autocorrelation function
\begin{equation}
C_Q(t)\equiv \frac{\langle Q(t)Q(0)\rangle-\langle Q \rangle ^2
}{\langle  Q^2 \rangle- \langle Q \rangle ^2}.
\end{equation}
In Fig.~\ref{fig:qpm-time}(a) we show $C_Q(t)$ for different
temperatures. The decay of $C_Q(t)$ is reminiscent of the decay of the
self intermediate scattering function. The long time behavior of
$C_Q(t)$ is exponential at high $T$, but at low $T$ can be fit with a
stretched exponential $\exp[-(t/\tau)^\beta]$, where $0<\beta<1$. We
define the correlation time $\tau_Q$ as the time required for $C_Q(t)$
to decay by a factor $e$. Figure~\ref{fig:qpm-time}(b) shows the values
of $\tau_Q$ as function of $1/T$ on an Arrhenius plot. The behavior of
$\tau_Q$ is non-Arrehenius at high temperatures and can be fit by a
power law $(T-T_{\rm MCT})^{-\gamma}$ where $T_{\rm MCT}\approx235$ is the
mode coupling temperature \cite{xuPNAS,gotze}. At low $T$, $\tau_Q$
deviates from the power-law fit and becomes Arrhenius. This crossover in
relaxation of local orientational order occurs near $T_W$.

\begin{figure}
\begin{center}
\includegraphics[width=8cm]{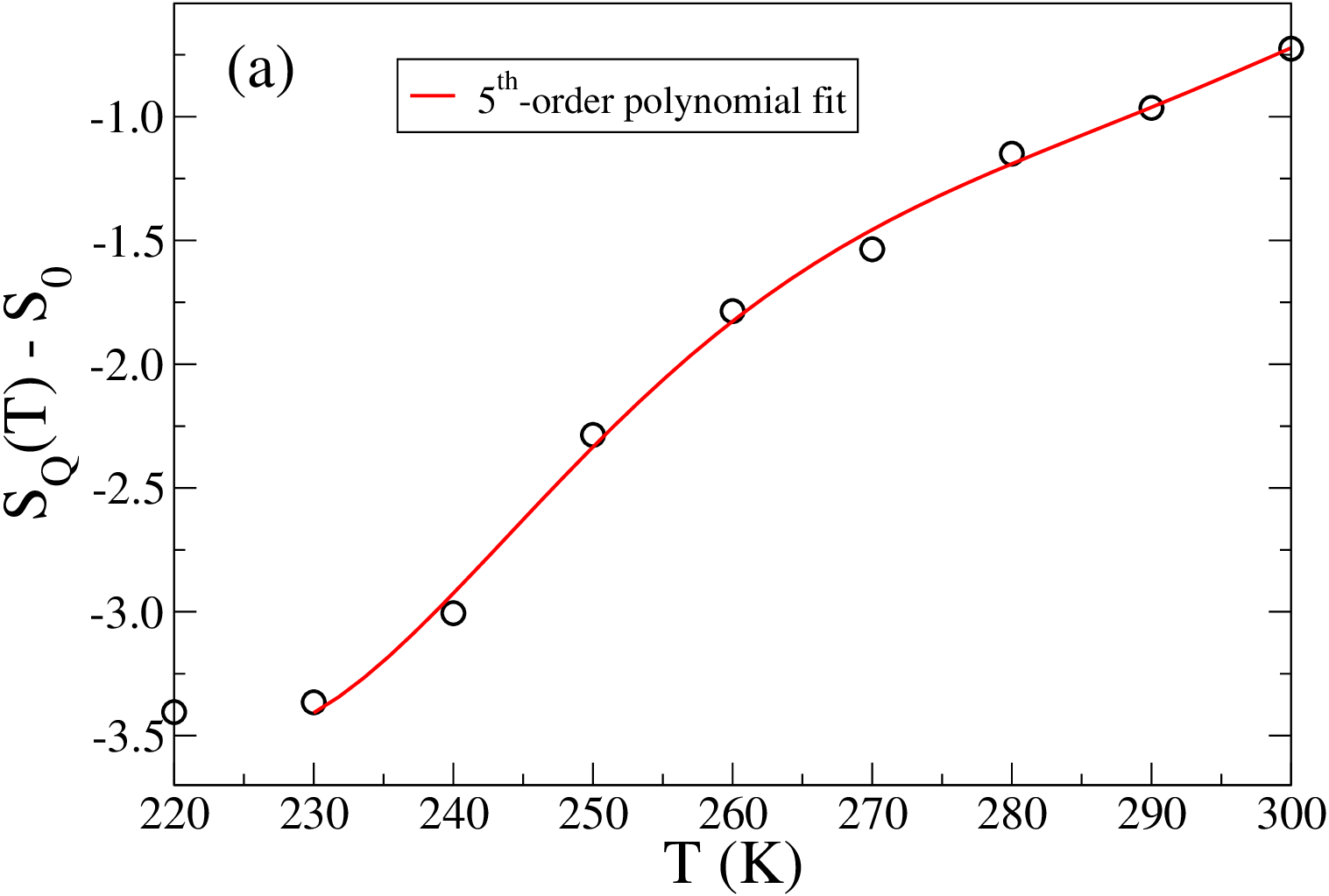}
\includegraphics[width=8cm]{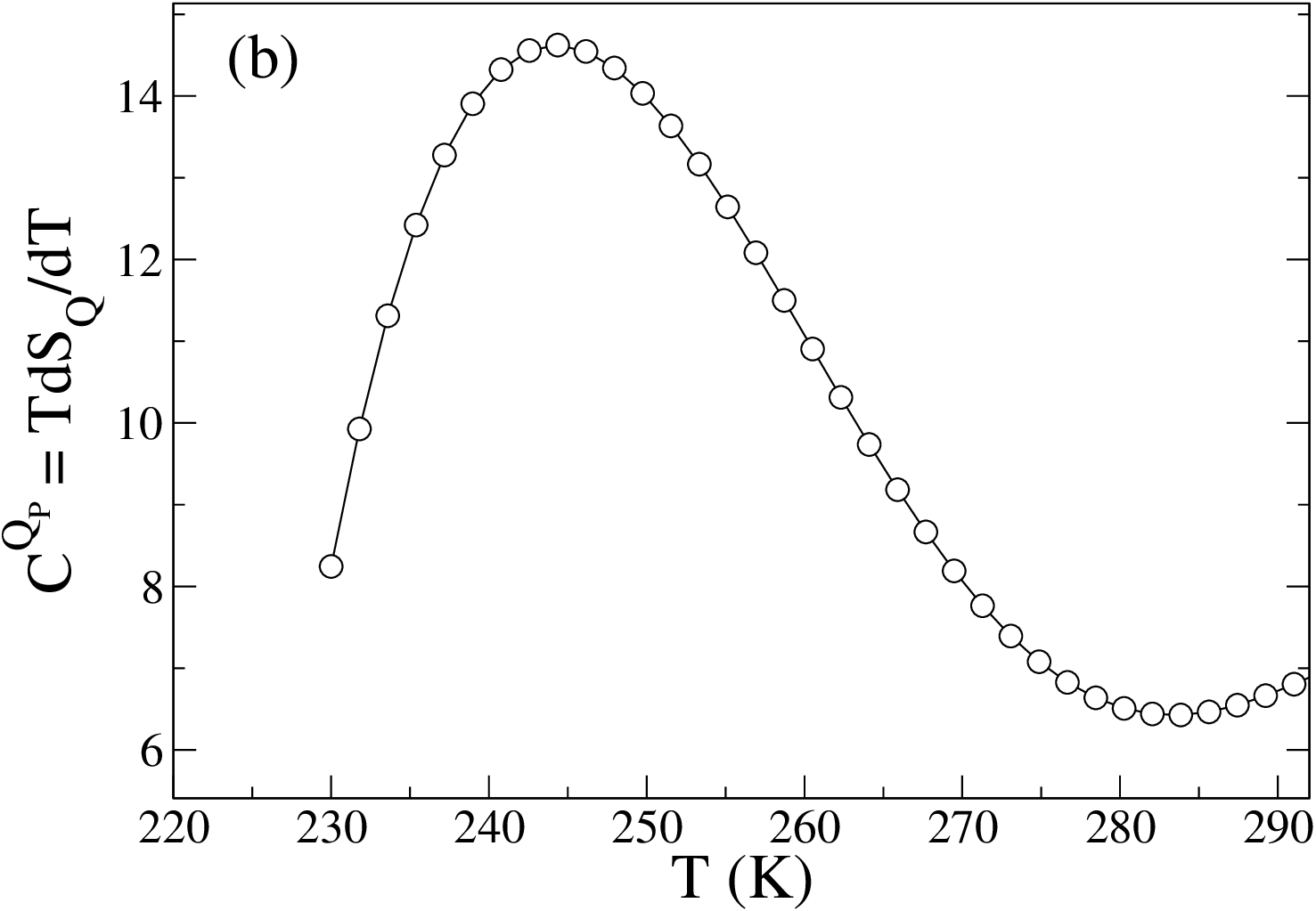}
\end{center}
\caption{(color online) Temperature dependence of (a) orientational
 entropy $S_Q(T)$, defined in Eq.~(\ref{eq:ln}), and (b) orientational
 specific heat $C_P^Q=T\left(\partial S_Q/dT\right)_P$, which has a
 maximum around the same temperature, $T\approx 246$~K~$\approx T_W$,
 where the total specific heat $C_P$ has a maximum. Solid line in (a)
 is a $5^{th}$-order polynomial fit through the data which is used to
 generate (b)}
\label{fig:sq}
\end{figure}

Since $Q$ measures the local orientational order, it must contribute to
the entropy of the system.  We next derive an expression for this
``orientational entropy'' $S(Q_1,Q_2,...,Q_N)$, which we define to be
the logarithm of the number of states corresponding to the interval
between $(Q_1,Q_2,....,Q_N)$ and $(Q_1+\Delta Q_1,Q_2+\Delta Q_2, ...,
Q_N + \Delta Q_N)$.  According to Eq.~(\ref{eq:qk}), $8(1-Q_k)/3=const$
defines a surface of a six-dimensional hypersphere of radius
$\sqrt{8(1-Q_k)/3}$ in the space defined by the six tetrahedral angles
$\psi_{ikj}$ of Eq.~(\ref{eq:qk}). Hence we assume that the number of
states between $Q_k$ and $Q_k + \Delta Q_k$ of molecule $k$ scales as
$\left (1-Q_k \right )^{5/2}\Delta Q_k$.  We assume that the order
parameters of each molecule is independent, an assumption justified by
the small value of $C_Q(r)$ studied. Thus we can define the number of
states $\Omega (Q_1,Q_2,...., Q_N)$ in the interval between
$(Q_1,Q_2,....Q_N)$ and $(Q_1+\Delta Q_1,Q_2+\Delta Q_2, ..., Q_N
+\Delta Q_N)$ as the product
\begin{equation}
\Omega (Q_1,Q_2, Q_3, ...., Q_N)
\equiv N\Omega_{0}\displaystyle\prod_{k=1}^{N}(1-Q_k)^{\frac{5}{2}},
\label{eq:ln0}
\end{equation}
where $\Omega_{0} = const$.  Hence the orientational entropy of the
entire system is given by
\begin{equation}
S(Q_1,Q_2,Q_3,...,Q_N) \equiv N S_{0} + {\rm ln}
\displaystyle\prod_{k=1}^{N}(1-Q_k)^{\frac{5}{2}} \equiv N S_{0}+
\frac{5}{2} \displaystyle\sum_{k=1}^{N} {\rm ln} (1-Q_k).
\label{eq:ln}
\end{equation}
where $S_0=\ln\Omega_0$. If $P(Q,T)$ is the distribution of $Q$ at a
given temperature $T$, then the orientational entropy $S_Q(T)$ per
particle at temperature $T$ can be written as
\begin{equation}
S_Q(T) \equiv S_{0} +  \frac{5}{2}
\displaystyle\int_{Q_{\rm min}}^{Q_{\rm max}}
{\rm ln} (1-Q) P(Q,T) dQ.
\label{eq:SQ}
\end{equation}
In Fig.~\ref{fig:sq} (a), we show that $S_Q(T)-S_{0}$ decreases with
decreasing temperature as expected.

We further define a measure of ``orientational specific heat'' as
$C_P^Q(T)\equiv T(\partial S_Q(T)/\partial T)_P$. Figure~\ref{fig:sq}(b)
shows the temperature dependence of $\displaystyle C_{P}^{Q}(T)$ for
$P=1$~atm. $\displaystyle C_{P}^{Q}(T)$ shows a maximum at $T \approx
245$K $\approx T_W$, where the total specific heat $C_P$ has a maximum
\cite{xuPNAS,kumarPNAS07}, suggesting that the fluctuations in
orientational order reach a maximum at $T_W(P)$.

To relate the orientational entropy $S_Q(T)$ to the orientational
relaxation time $\tau_Q(T)$ associated with the local tetrahedral
ordering, we propose the following generalization of the Adam-Gibbs
relation between the translational relaxation time and configurational
entropy \cite{AG,giovamb03},
\begin{equation}
\tau_Q(T)=\tau_Q(0)\exp[A/TS_Q(T)],
\label{eq:ag}
\end{equation}
where $\tau_Q(0)$ is the orientational relaxation time at very large
$T$, and $A$ is a parameter playing the role of activation
energy. Accordingly we calculate [Fig.~\ref{fig:qpm-time}(b)]
$\tau_Q(T)$ from $S_Q(T)$ using Eq.~(\ref{eq:ag}) with three free
parameters: $A$, $\tau(0)$, and $S_{0}$. We find that $S_{0}\approx
4.33$.

Figure~\ref{fig:qpm-time}(b) is an Arrhenius plot of $\tau_Q$,
calculated from Eq.~(\ref{eq:ag}). The temperature dependence of
$\tau_Q$ is different at low and high temperatures, changing from
non-Arrhenius (a T-dependent slope on the Arrhenius plot) which can be
fit by a power law at high temperatures to Arrhenius (a constant slope)
at low temperatures \cite{Liu04,xuPNAS,kumarPNAS07}.

In summary, we have studied the space and time correlations of local
tetrahedral order, presumably related to local orientational
heterogeneities.  We find that the spatial correlation of the the local
tetrahedral order is anticorrelated for the molecules separated by
3.2\AA~ at high temperatures.  This negative correlation changes to
positive correlation upon decreasing $T$ below the Widom temperature
$T_W(P)$.  Further, we define a measure of the orientational entropy
$S_Q$ and find that it well describes orientational relaxation using the
Adam-Gibbs relation.

\bigskip\bigskip

\noindent We thank P. H. Poole, S. Sastry, and F. W. Starr for fruitful
discussions and the NSF Chemistry Division for support.  S.V.B.
acknowledges the partial support of this research through the
Dr. Bernard W. Gamson Computational Science Center at Yeshiva College

\newpage

\newpage

\end{document}